\def\Fp{F_{\rm p}}
\def\Cp{C_{\rm p}}
\def\F2{F_{\rm 2}}
\def\Em{E_{\rm m}}
\def\Bm{B_{\rm m}}
\def\me{m_{\rm e}}
\def\qe{q_{\rm e}}
\def\r0{r_{\rm 0}}
\def\e0{\epsilon_{\rm 0}}
\def\u0{\mu_{\rm 0}}
\def\V0{V_0}
\def\Ug{U_{\rm g}}
\def\Ue{U_{\rm e}}
\def\Ev{\vec{E}}
\def\Av{\vec{A}}
\def\Er{E_{\rm r}}
\def\Eth{E_\theta}
\def\Ephi{E_\phi}

\def\rv{\vec{r}}
\def\vvr{\vec{v}_{\rm r}}
\def\rvr{\vec{r}_{\rm r}}
\def\rr{r_{\rm r}}
\def\rur{\vec{\hat{r}_{\rm r}}}

\documentclass{aa1}

\journalname{Astronomy and Astrophysics}
\begin{document}

\title{Could the pulsars actually be oscillators?}
\author{Gary Osborn\inst{}}

\institute{Electro--Chemical Devices, 23665 Via Del Rio, Yorba Linda CA
  92887 USA, \email{gary@s-4.com}
  }
\date{Received: date / Revised version: date / Version: 27 February 2001}
%
%
\abstract{
A simplified form of a nonsymmetric metric is used to develop the
coupling of the electrostatic and gravitational fields, which only
occurs in dynamic solutions.  The coupling results in a
resonance in an object near the Schwarzschild radius.  The resonator is
unique in that the frequency decreases with radius, reaching a lower
limit near 0.4 Hz at the gravitational limit.  The limiting frequency
does not depend on the mass or any other parameter.  It is a constant of
physics.  Gravitational contraction supplies energy to the resonator,
causing it to break into oscillation.  The solution is in reasonable
agreement with the properties of the mainstream pulsars.  An
observational test to determine if the pulsars are oscillators has never
been performed, and an observational technique is developed for
distinguishing between the rotational and oscillatory cases.
}

\maketitle

\section{Introduction}

In the most general case mathematically possible the distance between
two infinitesimally spaced points has both symmetric and antisymmetric
components.  It is the conventional conclusion that the antisymmetric
terms do not occur in nature, but there is the consideration that the
general theory is not a theory of electricity.  The Maxwell equations
are accommodated by the theory; they are not obtainable from the
asymptotic limits of it.

There have been many investigations into the possibility that the
antisymmetric terms play a role \cite{damour}, but none of them has led to an
established theory.  It is suggested that some of the difficulty is due
to the lack of observational data of an appropriate form, and that
pulsar observations may be able to supply it.

The existing nonsymmetric
gravitational theories are not well enough developed that they can be
used to obtain a persuasive pulsar solution, while without observational
evidence that the pulsars are oscillators there is no compelling need to
develop them.  Progress in such areas has traditionally been made by a
series of small steps, with observational data leading the way.
The theoretical and observational consequences of
discovering that the pulsars are oscillators are so enormous that a
further analysis of the observational data seems justified.

The effort required in evaluating the data is minimal.  It is simply
that if the pulsars are oscillators then the bands of drifting subpulses
will tend to synchronize with the main pulse, while there is no
possibility of synchronism in an isolated rotating pulsar.  The test is
simple and effective in distinguishing between the two models.  There
are unfortunately no existing published data that are of an appropriate
form for resolving the question, with that being due to the widespread
but unvalidated belief that the pulsars are rapidly rotating neutron
stars.

A fully developed derivation of the pulsar solution will be a formidable
exercise, since the symmetric and antisymmetric terms uncouple in
linearized solutions (\cite{moffat}).  There appears to
be a simpler way of obtaining the essential characteristics of the
solution.  The method is not very accurate, but the solution obtained
should be sufficient for the initial evaluations, and if the pulsars are
found to be oscillators then a substantial amount of theoretical work in
the literature will be brought to bear on the problem.

It might seem that after more than 30 years of pulsar observation that
there could be no question of the outcome.  The problem is harder than
that, because point--like oscillating and rotating radiators are not
distinguishable in any elementary way.  One reason that oscillatory
pulsar models have not received much attention is that the resonant
frequency must decrease as the radius decreases if gravitational
contraction is to supply the energy.  The oscillator developed here is
probably the only one in existence with that characteristic.

\section[]{The electrical scaling relationships}

The behavior of light rays in the Schwarzschild solution can be
represented by a refractive index model \cite{felice}.  The solution is only
valid from the perspective of a distant observer, which corresponds to
the perspective of a potential theory.  The speed of light is rarely $c$
in such a representation.  The perspective is that of the space of
the potentials, and the potentials do not correspond directly to real space.

Beginning with a simplified approach, and considering only the
electrostatic and gravitational fields at first, simulate the
antisymmetry by starting with two synchronized clocks, then displace one
of the clocks in time.  The indicated propagation velocity in one
direction is now greater than $c$, while the indicated velocity in the
other direction is less than $c$.  One part is symmetric and one part is
antisymmetric, and neither part is $c$.  (However, for a local observer
only, the symmetric part is $c$ even in the gravitational field, and
indeed the symmetric part can be interpreted as always being $c$ in a
non-flat space.  The propagation velocity in these equations does not have
the same meaning, as it is evaluated from the perspective of a distant observer
in a fictitious flat coordinate system.)

A field of displaced clocks represents a scalar potential.  If the
potential is of electrical significance then the only simple possibility
is that the gradient is the $E$ field.  More generally, the 4-potential is
a 4-vector, so it transforms in the same way as the coordinates.  The
mathematical correspondence between the scalar potential of the Maxwell
equations and displacements in time is extremely close.

The scaling relationships will be needed.  \cite{shabad}
derived a critical magnetic field strength where a resonance condition
increases the efficiency of gamma ray to electron--positron pair
conversion.  Multiplying that equation by $b$, with $b$ being a small
undetermined correction due to the different meaning of these
relationships, the equation becomes
\[
\Bm = b c^2 \me^2/(\hbar \qe) = b \times 4.41 \times 10^9\; {\rm tesla,}
\]
with $\hbar = h/(2 \pi)$, $h$ Planck's constant, $\qe$ charge of electron,
$\me$ mass of electron.

The $E$ and $B$ fields carry equal energies in electromagnetic radiation,
implying the equation $\frac {1}{2} \e0 \Em^2 = \frac{1}{2} \Bm^2/\u0$.
Solving for $\Em$ and applying the identity $c = (\e0 \u0)^{-\frac{1}{2}}$
\[
\Em = b c^3 \me^2/(\hbar \qe) =
b \times 1.32 \times 10^{18}\;{\rm V\: m^{-1}}.
\]

The Dirac equations predict that an intense enough $E$ field produces
electron--positron pairs so prolifically that further increases are not
possible, which represents a field limit.  The limit is not precisely
defined, but it is of a magnitude similar to $\Em$.  A static magnetic
field cannot decay into single electron--positron pairs, but decay into
multiple particles might be possible \cite{calucci}.

Returning to the antisymmetric scaling relationship, if the scalar
potential represents a displacement in time then a certain $E$ field
exists at which a pulse transmitted from one capacitor plate to the
other arrives at the same time it was transmitted.  At higher voltages
it arrives before it was transmitted, which is an unlikely circumstance.
A field limit is implied, and the field limit establishes the scaling
relationship.  The field limit of the Dirac equations is a likely
identification.  Then, with the sign of $\xi$ being undetermined,
\[
1/(\Em c) = \xi = \hbar \qe/(b c^4 \me^2)
= b^{-1} \times 2.52 \times 10^{-27} {\rm \: s\: V^{-1}}.
\]

A field of displaced times can be transformed.  One of the ways of
deriving the Li\'{e}nard--Wiechert retardation equations consists
of transforming to the frame of reference of a point charge.  The
solution represents a static inverse square law field in that
system, which is transformed back to the original frame of reference.
Reviewing the derivation \cite{morse} shows that if the scalar potential is
replaced by a displacement in time the only effect is to change the
system of units, with the solution being

\begin{eqnarray*}
\Delta \rv &=& \xi q \vvr /[4 \pi \e0 \rr (1 - \rur \cdot \vvr /c)] \\
\Delta t &=& \xi q/[4 \pi \e0 \rr (1 - \rur \cdot \vvr /c)].
\end{eqnarray*}

The solution is the Li\'{e}nard--Wiechert
retardation equations in a system of units in which the
scalar potential represents a displacement in time and the vector potential
represents a displacement in space.  $\vvr$ is the retarded velocity and
$\rvr$ is the vector from the particle at the retarded position to
the field point.  All solutions to the retardation equations are also solutions
to the Maxwell equations in potential form, from which it follows that
those equations are also compatible with this system of units.
But does this system have a special physical significance?  It is
doubtful that the question can be answered by purely mathematical
methods.  Further, the electrostatic force is a strong force and
the displacements are extremely small, making laboratory evaluation
of the equations difficult or impossible.  But if the potentials
do represent metrical displacements then, while the Maxwell equations
are linear, the displacements will affect light rays in ways not
represented by the Maxwell equations, with the result being
that the full system of equations becomes self--interacting and
nonlinear.  The full system is therefore not the Maxwell equations, but
is a system that is not yet known.  It is nevertheless sometimes
possible to make meaningful calculations in an incompletely understood system
from energy and symmetry considerations.

\section[]{The electrograv resonance}

\subsection[]{The space--time coupling}

The analysis will utilize the quasi--static form of the Maxwell
equations, which is often satisfactory for low frequency problems.  For
the oldest mainstream pulsars the wavelength exceeds the diameter by a
factor of $10^5$, justifying the simplification.  The coupling of the
electrostatic and gravitational fields varies with frequency as $\omega^1$,
and the time--varying part of the magnetic field also varies as $\omega^1$ in
these solutions, so the magnetic--gravitational coupling varies as
$\omega^2$, and it can be neglected at low frequencies.

Begin with two conductive masses connected by a wire.  Cut the wire at
the midpoint and insert an oscillator.  Mount a clock on each mass.
Also place a reference clock near the midpoint.  Measure time by reading
the clock faces.  The magnitude of a potential is not locally
detectable, so an observer riding with the mass on its journey in time
will conclude that the local system behaves in an entirely conventional
way. $\partial \psi/\partial t$ behaves like the expansion factor of the
space part, and the expansion factor is not locally detectable either.  The mass
emits a gravitational flux which moves radially outward at $c$.  As the
voltage increases with time (with an undetermined polarity) the clock
moves into the future relative to the reference clock.  The time
interval of the moving clock maps into a shorter time interval at the
reference clock.  An observer at the reference point therefore perceives
a greater time--integrated flux from the source.  So, if the
scalar potential does represent a displacement in time, then a changing
electrostatic field modulates the intensity of the gravitational field.

The voltage on one mass is $\frac{1}{2} \V0 \sin \omega t$.  The voltage on
the other mass is $-\frac{1}{2} \V0 \sin \omega t$.  The distance between the
masses is $d$, and their static gravitational binding energy is
$-G m_1 m_2/d$.  The magnitude of the one way modulation is
$\xi {\rm d}V/{\rm d}t$, and the modulated field strength enters
into the energy equation in the same way as the
magnitude of either mass, so from the preceding considerations the
energy modulation is
\[
(1 - \V0 \xi \omega \cos \omega t) (1 + \V0 \xi \omega \cos \omega t).
\]
In being proportional to ${\rm d}V/{\rm d}t$ the coupling seems to
be linear, and in some sense it is, but in the overall system the coupling
is associated with nonlinearity.  That is more clear if the same method
of analysis is applied to two charges, where the interactions produce fields
quadratic in source strength.  The significance is that in a fully
developed derivation the electrostatic--gravitational coupling will be
lost if the equations are linearized, because the coupling is a
space--time cross term that is in the same order as the quadratic terms.
The full system of equations is nonlinear, and it will be
difficult to obtain the complete solution.

Proceeding similarly, if the gravitational flux from a parent body is
enhanced in propagating to a test mass then the flux from the test mass
propagates to the parent body along a path where the $E$ field is of the
opposite sign.  The acceleration at the outer surface of a thin
spherical mass shell is $G m/r^2$.  The acceleration at the inner
surface is zero, so the average acceleration is $\frac{1}{2} G m/r^2$.
Beginning with a shell of infinite radius and integrating force times
distance leads to the gravitational potential energy, $-\frac{1}{2} G m^2/\r0$.
The gravitational energy can also be computed by integrating $-a^2/(8 \pi
G)$ over all space, where $a$ is the acceleration.  The following
calculations proceed by multiplying the energy density by the modulation
factor, then integrating.  The pulsar is assumed to be a good electrical
conductor, so the $E$ field does not penetrate to the interior region.

Without supposing that the appropriate pulsar spherical harmonics are
yet known, begin with the lowest order Legendre polynomial that
conserves charge, $P_1$.  The spherical harmonics represent the charge
density on the surface of the sphere.  The radial dependency must be
computed.  The static solution is multiplied by $\sin \omega t$.  The
effects of retardation are not considered so the result is not an exact
solution to the Maxwell equations, but it is quite close when the
frequency is low.  Then
\[
\psi = \V0 \r0^2 \sin \omega t \cos \theta/r^2.
\]
Neglecting the vector potential $\Av$ at low frequencies, the $\Ev$ field
is $-\nabla \psi$.
\begin{eqnarray*}
\Er &=& 2 \V0 \r0^2 \sin \omega t \cos \theta/r^3\\
\Eth &=& \V0 \r0^2 \sin \omega t \sin \theta/r^3\\
\Ephi&=&0 \end{eqnarray*}

The energy density is $\frac{1}{2} \e0 \Ev \cdot \Ev$.
\[
2 \V0^2 \e0 \r0^4 \sin^2 \omega t/r^6 -\frac{3}{2} \V0^2 \e0 \r0^4 \sin^2
\theta \sin^2 \omega t/r^6
\]
Integrating over the exterior region,
\[
\Ue = \frac{2}{3} \pi \V0^2 \e0 \r0 (1 - \cos 2 \omega t).
\]
The fringing capacitance between the northern and southern hemispheres
forms a capacitor.  The equation gives the energy stored in the
capacitor.  The gravitational inductance of the conductive sphere
prevents the capacitor from being short--circuited.

The gravitational energy density is $-a^2/(8 \pi G)$, with $a = G m/r^2$.
Multiplying by the modulation factor leads to
\[
-G m^2/(8 \pi r^4) +G \V0^2 m^2 \r0^4 \omega^2 \xi^2 \cos^2 \theta \cos^2
\omega t/(8 \pi r^8),
\]
which is integrated over the exterior region.
\[
\Ug = -G m^2/(2 \r0) + (1 + \cos 2 \omega t) G \xi^2 \V0^2 m^2 \omega^2
/(60 \r0)
\]
At time $t=0$ the voltage is zero but it is changing at its maximum
rate.  Viewing the equation for the total energy as representing the sum
of a constant negative binding energy and a positive periodic energy
flow shows that the positive energy is at a maximum when the electrical
potential difference is zero.

The energy stored in an inductor is $\frac {1}{2} L I^2$, where $I$ is the
current.  When resonated with a capacitor the energy stored in the
inductor is also at a maximum when the voltage across it is zero.  Thus,
at any given frequency, the coupling of the electrostatic and gravitational
fields looks like an inductor.  The frequency dependence is that of a
capacitor, but it is the phase relationships that matter in computing
the resonant frequency.  For dense objects the gravitational inductance
is far greater than the Maxwellian inductance, so the latter will be
neglected.

Now connect an oscillator between the polar regions of the conductive
mass.  The energy equation shows that, when averaged over many cycles,
the magnitude of the gravitational potential energy is less than before.
The modulation weakens the gravitational field.  The object is not in
equilibrium, so its resilience will increase the radius.  When the
oscillator is turned off the object will return to its original radius,
and the electrical energy transferred to the system could be recovered
during the return.

Next, allow the radius to reach equilibrium with the oscillator on, then
slowly compress the mass.  The oscillator did work on the system in
increasing the radius, so the system must do work on the oscillator when
the mass is compressed.  Gravitational contraction can power the
oscillator.

At time $t=0$ all of the resonator energy is stored in the inductor.  At
the $90^{\circ}$ point on the cycle all of the energy has been transferred
to the capacitor, so the time--dependent portion of $\Ug$ at the first
time must be equal to $\Ue$ at the second time.  The two equations are
solved for $\omega$, then divided by $2 \pi$ to obtain the frequency in Hz.
The solution is then reparameterized by the substitution $\r0 = 2 k G m/c^2$.
\[
f = (40 \e0 G/\pi)^{1/2} k/(c^2 \xi)
\]
The parameter $k$ represents the ratio of the actual radius to the
Schwarzschild radius.  The solution evaluates numerically to $0.383 k b$
Hz.  Proceeding similarly, the $P_2$ solution is $0.555 k b$.  The $P_3$
solution is $0.726 k b$ and the spherical harmonic $\sin 2 \phi \sin \theta$
comes in at $0.557 k b$.

\subsection[]{Limitations}

The gravitational redshift will significantly lower the computed
frequency of the oldest pulsars.  The calculation is Newtonian, so there
are several other inaccuracies near the Schwarzschild radius.

Self--initiated oscillation modes probably cannot evolve on the surface
of a sphere without the preferred direction specified by a pre--existing
magnetic field.  The nonlinear coupling to the static field was not
included in these calculations, but its stabilizing influence is
probably essential.

The millisecond pulsars are outside the range of validity of the low
frequency approximations utilized if they are of normal mass, and
observational data indicate that they are \cite{callanan}.
There may be a second solution representing the coupling of the gravitational
and magnetic fields, with the $E$ field playing only an auxiliary role.
This kind of duality exists in the Maxwell equations, and it could carry
over into the gravitational solutions.  The difficulty in accommodating
the millisecond pulsars with this solution encourages attempts to find a
second solution.  The millisecond pulsars are not just ordinary pulsars
that operate at a higher frequency.  There are essential differences
between the two pulsar classes.

The oscillation amplitude will build to the point where some nonlinear
limiting process occurs, and that point is probably much less than $\Em$.
For example, a field of only $10^5 {\rm V m^{-1}}$ would be sufficient to
levitate a proton.  This field is weak even by laboratory standards, and
at some point it may become possible for surface charges to be swept
away by the $E$ field.  The mass ejection would become so large at that
point that further increases in oscillation amplitude would not be
possible.  If that happens then the pulsar pulses would be associated
with the peaks of the sine wave, which reach some critical threshold of
charged particle ejection.

\subsection[]{The spindown power}

The gravitational potential energy of a solid sphere, including the
internal energy, is $U = -\frac {3}{5} G m^2/\r0$.  The internal energy is
1/6 of the total.  Substituting $\r0 = k(t) 2 G m/c^2$ to parameterize by the
Schwarzschild radius yields $U = -\frac {3}{10} m c^2/k(t)$.  The outbound
energy flow is $-{\rm d}U/{\rm d}t = -\frac{3}{10} m c^2
({\rm d}k/{\rm d}t)/k^2$.  The frequency is
proportional to $k$, so the fractional frequency change per second is
$({\rm d}k/{\rm d}t)/k = ({\rm d}f/{\rm d}t)/f =
-({\rm d}P/{\rm d}t)/P$, where $P$ is the pulsar period.
Then, neglecting the mass lost by the equation $E=m c^2$, the energy
flow is
\[
p = 3 m c^2/(10 k P) \frac{{\rm d}P}{{\rm d}t}.
\]
This calculation includes neutrino radiation and the kinetic energy of
the ejected mass, except that the equation is not valid when the pulsar
ejects its own mass.  It overestimates the power output in that case.

Assuming $1.4 {\rm M}_\odot$ $(2.78 \times 10^{30}$ kg), a spindown rate
of one part in $10^{15}$ per second, and $k=3$, the equation evaluates to
$2.5 \times 10^{31}$ watts.  For $b=1$ and the $P_1$ mode the frequency
is 1.2 Hz and the radius is 12 km.

\section[]{Observational tests}

\subsection[]{Aligned rotors}

Charged particles can only escape through the magnetic field in the polar
regions.  Recent \emph{Chandra}  X-ray images of the Crab \cite{weisskopf}
and Vela \cite{helfand} pulsars show jets that appear to
be parallel to both the magnetic and spin axes, since if the axes
were not aligned the jets would take the form of cones.
A rotating pulsar becomes disabled when the axes are aligned.

The magnetic force between the poles of a magnetized sphere is
attractive.  The force at the equator is repulsive, so the magnetic
field causes an equatorial bulge.  Centrifugal force also causes an
equatorial bulge.  The energetically preferred orientation of the
magnetic field is therefore parallel to the spin axis.

\subsection[]{The low frequency cutoff}

The oldest radio pulsars drop out at about 0.25 Hz.
\cite{baring} have proposed that the pulsars become radio--quiet because of
photon splitting in the intense magnetic field.  The cutoff point is
the same as the equation for $\Bm$ with $b=1$.  There are several isolated
X-ray pulsars in the 0.1 to 0.2 Hz range, leaving open the possibility
that the pulsars continue to oscillate after they become radio--quiet.
The gamma ray bursts sometimes exhibit a light curve resembling the
exponential decay of a resonant system with a frequency of 0.1 to 0.2
Hz.  In the case of the 1998 August 27 burst from SGR 1900+14
the period of the decaying light curve was the same as the
pulsar period \cite{hurley}, tending to establish a connection between the two
phenomena.  The lowest frequency isolated pulsar known has a frequency
of 0.085 Hz \cite{vasisht}.  In view of the errors inherent
in a Newtonian solution near the Schwarzschild radius, and also
considering the neglect of the gravitational redshift, these observed
limiting relationships are in reasonable agreement with the predicted
cutoff frequency.  The pulse frequency could be at the second harmonic
of the oscillation frequency, in which case the discrepancy is greater.

The accretion--powered X-ray pulsars, which are found in binary systems,
have periods up to 1400 seconds \cite{bildsten}, and are known to
be routinely spun up by accretion.  Pulsars necessarily do rotate, and
the oscillatory model does not apply to the long period accretion--powered
pulsars, except that detecting sustained high frequency
oscillation in one of these systems would constitute an excellent
observational test.

\subsection[]{The energy deficit}

When applied to the 7.47 second isolated X-ray pulsar SGR 1806-20 the
oscillatory model overestimates the observed power output of the full
system, including the nebula, by a factor of $10^5$.  The rotational
pulsar model fares no better, as it underestimates the total power
output by a factor of $10^3$ if the pulsar radius is 10 km.  The most
commonly accepted explanation for the energy deficit is that the pulsar
is powered by the decay of an unusually intense magnetic field rather
than by rotational kinetic energy \cite{kouveliotou}.  The
magnetar field estimates exceed the critical field of the Dirac
equations by up to a factor of 20 \cite{gotthelf}.
That may be possible, but the field strength estimates do not appear to
be reliable.  The spindown rate of SGR 1900+14 doubled for
80 days following the
August 1998 outburst, suggesting that magnetic braking is not the
spindown mechanism for this pulsar, since the field strength would have
had to increase by a factor of two in a short time, which is not
plausible \cite{marsden}.

The oscillatory power calculation is based only on basic energy
relationships, so the energy discrepancy in this model may be due to
mass ejection.  In this model the pulsar has an intense $E$ field so
when infalling material becomes ionized half the particles are ejected
from the system.  The other half collides with the pulsar, creating more
charged particles.  At high energies a relatively small mass flow could
account for the computed energy transfer, but a more likely explanation
is that the equation badly overestimates the actual power output if the
pulsar ejects its own mass, which can happen in an electrostatic field.
The computed power output must therefore be taken as only an upper limit
until observational data on the pulsar's mass flow become available.
Infrared observations show that the pulsar is surrounded by a dust cloud
\cite{smith}, which may be relevant to the energy flow equation.

The 0.085 Hz pulsar 1E1841-045 radiates $3.5 \times 10^{28}$ watts in X-rays
alone.  The 0.091 Hz pulsar 1RXS J170849.0-400910 is similar, with a
steady X-ray output of $1.2 \times 10^{29}$ watts \cite{israel}.  These
power levels cannot be supplied by the rotational kinetic energy of the
pulsar, and it is likely that there are other and still--unmeasured
energy flows to be considered.

\subsection[]{The second resonance}

There are an infinity of spherical resonance modes, and
it may be possible for more than one mode to oscillate
simultaneously.  Even if the other modes are not oscillatory,
the main pulse is narrow and the harmonic content high, so it
can excite the high frequency modes that are almost harmonically
related, and the effect is about the same.

The 0.54 Hz pulsar PSR 0826-34 exhibits drift bands over a wider region
than most, having 5 bands spaced over nearly $200^{\circ}$ of the pulsar
cycle \cite{biggs}.  The uniformity of the spacing implies that
there are other hidden bands.  The autocorrelation function shows the
band spacing to be $29^{\circ} \pm 2$, so there must be a total of
12 bands in $360^{\circ}$, implying the existence of a resonance at the
$12^{\rm th}$ harmonic of $\Fp$.  The bands wander about in unison,
but they do
not drift systematically, implying that the resonance is almost exactly
at a harmonic.  But in general the spherical modes are not harmonically
related, and in most drifters the drift rate from pulse to pulse
evidently gives an indication of the extent that the resonances differ
from an integer ratio.  Such high order spherical resonances are densely
spaced, and a mode that has a frequency which is spaced as closely as
possible to an integer times the pulsar frequency will automatically be
selected.  This coupling would not occur when the higher frequency is
not an exact harmonic if the main pulse always occurred at the same
position, but it doesn't.  The high frequency resonance has enough
reactive energy flow that it can either dump the main pulse prematurely
or defer it slightly, depending on the phasing requirements, which is
illustrated in figure 2 of \cite{vivekanand}.
In a variation of this scheme the higher frequency is at approximately
$n+\frac{1}{2}$ times the pulsar frequency and the crests of the waves coincide
on every other pulsar cycle.

It is surprising that after 30 years of study the first
Fourier transform of the subpulse drift pattern extending to frequencies
higher than 0.5 of pulsar frequency will not be published until the year 2001
\cite{deshpande00}.  It is unobvious that such high frequencies should
be of interest, but the data for the 0.91 Hz pulsar B0943+10 imply that
much is to be learned of that region.

Figure 3 of the reference shows the spectrum of 256-pulse
sequences.  There is a broad
series of peaks at around 37 times the pulsar frequency $\Fp$, suggesting that
there is a resonance at that frequency.
The signal is only visible for
about $20^{\circ}$ out of the $360^{\circ}$ of the pulsar cycle, which
causes an extensive sideband system.  The modulation index is so high
that it impossible to determine
which peak is the resonance itself.  However, each peak is narrow,
meaning that the phase coherence from one pulsar cycle to the next is
excellent.  The spectrum is interpretable as representing
a highly monochromatic signal, $\F2$, which is modulated by the pulsar
frequency.

The peaks are at frequencies of $n+0.5355 \pm .0003 \Fp$.
The second harmonic is at $2 (n + 0.5355) = (2 n + 1 + .071) \Fp$, and these
components are also present in the figure,
although the $2 n$ value is aliased by $\pm 1$.
The intermodulation products of $\F2$ and $\Fp$ produce signals with periods
much longer than the pulsar period.
There are $n + 0.5355$ cycles of the higher frequency in one pulsar cycle,
and in 28 cycles there are $28 n + 14.994 \pm .008$ cycles,
with n being about 36.  That is approximately an integral number of cycles
for both frequencies, so the system
repeats then.  The pulsar sometimes switches modes, with the drift pattern
becoming disorganized in the other mode.

The Fourier transform also shows a peak at $0.0697 \pm .0005 \Fp$,
implying a repeating pattern of about 14 cycles.  But since the
resonance is at approximately $(n+\frac{1}{2}) \Fp$, an even pulse
nearly matches odd pulses that are displaced by 13 and 15 cycles,
causing the peak to be at the second harmonic of the
actual repeat interval.  If even pulses are matched with even pulses
then the repeat interval is $28.7 \pm 0.2$ cycles,
which is not in agreement with the preceding analysis.  Another estimate from
the same paper places the peak at $0.0710 \pm 0.0007$, and a repeat interval
of 28 cycles is within the error bound in this case.  The uncertainty
is probably best resolved by folding on various intervals.

The data also show two other weaker sidebands at
about $0.46 \pm .027 \Fp$, and they imply a repeat interval in the
vicinity of $37 \Cp$.  The autocorrelation function shows a broad peak
in the 35--40 cycle range, again implying that the repeat interval is
not 28 cycles.  However, a similar autocorrelation peak is still there in
the ``Q'' mode of the pulsar, where the $36 \Fp$ spectral feature is missing,
so it is evidently not associated with the 28-cycle sequence.
Further, the ``B'' mode autocorrelation
shows a local maximum at a lag of 28 cycles, and the even--odd pattern
causes the correlation to be negative at 27 and 29 cycles, supporting
the interpretation that the broad peak is unrelated to the basic
repeat interval. The data in the graph were smoothed on an interval
of 5 cycles, which, because of its even--odd pattern, discriminates
against the true amplitude of the 28-cycle peak.
More than one spherical mode can be excited, and it is unlikely
that any given drift sequence is completely pure.  Figure 13 of
the reference shows some weak sidebands at around $24 \Fp$ in the
Q mode, suggesting that other resonances do play a role.

The sidebands at $n+0.5355 \Fp$ extend to negative values of $n$, with
the negative frequencies being folded into the positive region.
That results in another sideband system at $m+0.4645 \Fp$.  The folded sidebands
are weak, and are only significant at low frequencies.  They are visible
if the e-print version of the figure is electronically magnified.
The folded and unfolded sidebands combine to form low frequency sideband pairs
at $(n \pm 0.5355) \Fp$, which
represents a modulation with a period of $1/0.5355 = 1.867 \Cp$.  At higher
frequencies only the upper sideband is significant, but that still represents a
modulation.  Figure 8 of the reference
is folded on this interval, confirming that a modulation exists.
The upper sidebands at $n+0.5355 \Fp$ can be viewed equally well as
being lower sidebands at $n-0.4645 \Fp$, so another periodicity at $2.153 \Cp$
is indicated.  Within observational accuracy, there are an
integral number of both the 1.867 and $2.153 \Cp$ periods in 28 pulsar cycles.

If the region 0 to $0.5 \Fp$ is transformed, as is commonly done,
the lowest frequency unfolded sideband at $0.5355 \Fp$ is above the
Nyquist frequency, and is aliased
to $1-0.5355 = 0.4645 \Fp$, which is the same as one of the negative
frequency sidebands.  Thus the peak at this frequency is the sum
of aliased and folded sidebands.  The high order harmonics dominate
the aliasing in this case, and the folded sidebands are weak in that
region, so the feature is mostly an alias.

The 1975 data \cite{sieber} for the same pulsar place
the aliased/folded peak at $0.473 \pm .002 \Fp$, which then corresponds
to a resonance at $36.527 \Fp$ in that epoch. In the figure showing
the drifting pulses the dotted
line marked ``A'' is one of the drift lines for even pulses.
The resonance is at approximately
$(n+\frac{1}{2}) \Fp$, so the pulses at a given location occur on every
other cycle.  Assuming that the frequency of the second resonance is the
1975 value, in two pulsar cycles there are 73.054 cycles,
and the nearest even--even or odd--odd peak occurs 0.054 of a cycle early.
In one pulsar cycle the value is half that.  The figure shows
18 pulses, so the phase should advance by $17 \times 0.027 = 0.459$
cycle in that time.  From this value the location of the drift band for
odd pulses, spaced by 0.5 cycle, can be computed.  Carrying through
the construction shows that the computed band spacing is 68 percent
greater than the actual spacing, meaning that the calculation
contains a systematic error.  One possibility is that nonlinear interactions
systematically change the phase of the reactive energy flow at the second
resonance each time a pulse is emitted, and there is evidence that missing
pulses do upset the regularity of the drift pattern in other pulsars
\cite{unwin}.

The drifters are not well understood in any
theory, and the full mechanism appears to be fairly elaborate.  The
mechanism is also difficult to discern, since in most cases only a few
degrees of the full $360^{\circ}$ cycle are visible in the data, and a
single high order resonance, combined with a narrow main pulse, can
produce a repeating pattern with a period of many pulsar cycles.  But
regardless of the mechanisms, there will be a tendency for all the
components to progress in a syncopated lockstep if the pulsar is an
oscillator, while there is no known mechanism that could account for
synchronism in an isolated rotating pulsar.

\subsection[]{Subharmonic phase lock}

For many pulsars the drift rate varies across the window, which is
interpretable as a ``phase pulling'' effect.  Such effects are universal
in free running oscillators, and they can lead to phase lock when two
coupled oscillators are almost harmonically related.  If phase lock
tends to occur then there will be preferred phase relationships between
the repeating pattern and the mean pulse profile, and these preferred
locations can be discovered by statistical techniques.

The full repeat interval is usually not
the same as the $P_3$ interval of the literature.  (In this model
$P_3$ is the time required for the signal at the second resonance
to drift through one full cycle.)  When $P_3$
is not an integral multiple of the basic pulsar period the
repeat interval is a multiple of $P_3$, with the constraint
being that the total interval must span an integral number
of pulsar cycles.  This relationship
has no meaning for rotating pulsars, but it is a vital
consideration in developing statistical tests to determine if there
is a tendency for the pattern to phase lock.
The full repeat interval must be reasonably short if there is to be
any possibility of detecting lock.

If the tendency exists then for a full repeat interval of $n$ pulsar
cycles there will be $n$ preferred phases of the repeating pattern,
with the preferred phases being separated by multiples of $\Cp$.
This test will be difficult to apply,
since disturbances will cause cycle slips, and the slips will destroy the
long term coherence at the repeat interval.
Further, even if the pulses do not have preferred positions,
but are more visible at some phases of the pulsar cycle (which
is known to be the case), then in any statistical process they
will seem to have preferred positions.  These relationships will make
it difficult to determine if preferred positions do exist.

Figure 3b of \cite{hankins} shows the
two--dimensional autocorrelation function for pulsar B1918+19.  It
shows that the $P_3$ interval is very close to $4 \Cp$.
Most of the features shown in the figure are explainable by a resonance
at $18.25 \Fp$, which causes the phase to advance by 0.25 cycle per
pulsar cycle.  One cycle of $\F2$ corresponds to
$360/18.25 = 19.7^{\circ}$ of the pulsar cycle.
With these values it is possible to compute the coordinates
in the figure of a grid of horizontal, vertical, and diagonal lines.
The diagonal lines are the drift lines.
The intersections of the three sets of lines agree quite well with the
correlation peaks.  The construction is not accurate enough to reliably
determine that $n$ is 18, but the Fourier transform will show a sharp peak
at the correct value.

Each of the four pulses that make up a cycle is visible
in most of the drift bands, but
that does not of itself mean that
the pulses have preferred positions.  Suppose that the pulse train
is not coherent with the main pulse.  Then for any arbitrarily selected
reference phase in the autocorrelation calculation the pulses
will occasionally drift to
that position.  In a long record (which this isn't) there will be regions
where the pulses coincide with any selected reference phase, and
a systematic pattern will exist that looks essentially the same.
If the pattern of seemingly preferred positions follows shifts
in the reference phase then they are not preferred positions
at all.  But if preferred
positions do exist then the intensity of the pattern will depend
on the reference phase, with the result being that the high density
points do not follow shifts in the reference phase.  Thus
if phase lock were exact and the reference
phase were midway between two preferred positions then the pattern would
vanish.  It is therefore not possible to determine if phase lock is a
tendency with any one autocorrelation calculation.

The figure also shows horizontal bands where the pulses do not drift.
They coincide
with pulse number 2 of the 4-pulse sequence.
Pulse 2 is special by being precisely at the peak
of the mean profile.  The interactions are plausibly stronger at that
point, resulting a ringing of the second resonance, and with two
cycles of the ringing being visible after each occurrence of pulse
2.  Pulse 2 always occurs in the same position, so the ringing
bands do not drift.  Observe also that when pulse 2 is early/late
the non--drifting bands are also early/late.  One problem with the ringing
hypothesis is that
pulse 2 also tends to be preceded by a weaker pulse.  Other
mechanisms may be at work in the system, perhaps a third and weaker
resonance.  The high frequency Fourier transform would answer that question,
and the low frequency transform does show that the periodicities are
different for the preceding pulses.

There is another way of checking for phase lock in this system.
Simply Fourier transform the entire observing session as a single
record, without regard for mode changes.  If phase lock occurs then each
time the pulsar returns to the mode shown in the figure pulse number
2 will still coincide with the peak of the profile, but, relative
to the prior data, the pulse can be slipped by 0 to 3 pulsar
cycles, which will destroy the coherence of the $18.25 \Fp$ frequency.
However, the pulse is narrow and the harmonic content high, and
the fourth harmonic is at $73 \Fp$, so the phase
at that frequency is unaffected by cycle slips.  Then if phase
lock tends to occur there will be a relatively strong spectral
feature at 88.91 Hz, and the central peak will have a spectral
purity approaching that of the pulsar itself.

The pulsar has two other drift modes, as well as a mode where
the drift pattern is disorganized.  It occasionally switches
between the these modes.  In mode A the Fourier transform shows
the repeat interval to be about $5.9 \Cp$, which places the second
resonance at $18.17 \Fp$.  In mode C the repeat interval is about $2.5 \Cp$,
with the resonance being at $18.4 Fp$.  The autocorrelation
for all three modes is shown in the reference, and a second resonance
near $18 \Fp$ is commensurate with all three modes.

A repeat interval
of $2.5 \Cp$ results in an even--odd pattern in the correlation bands,
since in one band the pulse phases are at 0, 0.4, and 0.8 cycle
of the $\F2$ frequency, while in the next band they are at phases
0.2, 0.6, and 0 or 1.0.  The result is two sets of drift lines, with
the phase advancing by 0.4 cycle each $\Cp$ in one set, and being retarded by
0.2 cycle in the other set.  Both sets of lines are visible in
the autocorrelation.

Then all four modes are explainable by
an oscillation that wanders between 18.17 and $18.4 \Fp$, with
the random pattern occurring when the frequency is not near any of
the three frequencies with drifting subpulses.  The drifting of the
second resonance could be tracked by Fourier transforming short sections
of the data record.  If phase lock tends to occur then certain frequencies
will be preferred.

Figure 4 of the reference shows that the mean profile has a
mode--dependent fine structure, suggesting that the subpulses
have preferred positions.  It can also be that the subpulses
are incoherent but are just more visible at some phases.  If they do
have preferred positions then a coherent subharmonic
relationship exists.  One way of seeing the impossibility of a coherent
subharmonic process in an isolated rotationally powered pulsar is to apply
the equivalence principle to the problem.  The principle also seems
appropriate in computing the effective $E$ field near a rotating pulsar,
since any plasma in the immediate vicinity will be dragged along
by the rotating magnetic field.  There are no absolute points
of reference in the local system.

\subsection[]{The rotational signature}

If the pulsars are oscillators then any low frequency and highly periodic
change in the pulsar signature might represent the rotational period.
The rotation would cause a small variation in the pulse arrival time, but it
is probably better technique is to search for periodic changes in the
pulse profile.  Several precessional effects are known in orbiting
pulsars, so only isolated pulsars can be considered in looking for the
rotational signature unless the rotation rate is so high that precession
can be ruled out.

One series of observations of the Crab pulsar showed a weak 60-second
periodic intensity modulation in the optical region \cite{cadez},
but the result was not repeatable \cite{golden}.  The best
example of a second period in
an isolated pulsar is found in the radio pulsar PSR B1828-11, which
exhibits a persistent 1000-day variation in both the pulse arrival time
and the pulse shape.  The second period has been interpreted as
being due to free precession \cite{stairs}.  Free precession requires
that the mass distribution be skewed with respect to the axis of
rotation \cite{pandey}.  Current pulsar models
assume that the interior is a superfluid, with a neutron crust.  With a
surface gravity of $10^{11}$ g it may be difficult to justify enough
deviation from isopotential stratification to account for the
precession.  If the 1000-day period represents the rotational period
then that would mean that the pulsars preferentially shed angular
momentum at birth.

It has been suggested that a radio lobe associated with SGR 1806-20 is
rotating with a period of 10 years \cite{kouveliotou}.
Further, the light curves shown in the reference appear to have
undergone a phase inversion in 5 years.  It will be difficult to
distinguish between the two models this way if the period of rotation
actually is measured in years, which is unexpectedly long.  Another way
of distinguishing between the oscillatory and rotational models is to
look for phase jumps.  A phase jump is possible in a rotating pulsar,
but it requires that the angular velocity first change with one sign,
then after a short time change again with the other sign, which is
improbable.  On the other hand for an oscillating pulsar any major
disturbance would be expected to affect the frequency, and the frequency
shift will integrate to a phase change after the event.  High frequency
oscillators are more susceptible to phase jumps.

\subsection[]{Glitches}

The radius decreases steadily as the pulsar ages, so the neutron crust
must rupture and buckle at times.  The buckling will increase the
effective radius, increasing the frequency of resonance.  The signatures
of the oscillatory and rotational model glitches are very similar in
this respect.

There is another factor to be considered in evaluating
the glitch characteristics.  The pulse profile of SGR 1900+14 changed at
the time of the August 1998 outburst.  The change persisted for 1.5
years and showed no signs of recovery \cite{woods}.
Some radio pulsars with drifting pulses
undergo similar profile changes.  They presumably represent a change in
the preferred scheme of phase locked frequency ratios, and occasional
readjustments will be necessary due to random disturbances, and
systematically so as the Schwarzschild radius is approached.  The
readjustment will affect both the basic pulsar frequency and the
spindown rate.  The possibility can be considered that a major and
sudden reshuffling of the frequency ratios is the cause of the glitches,
in which case they will be accompanied by profile changes.

\section[]{Discussion}

These equations predict that young pulsars are much larger than is
currently thought, and observational or theoretical data may exist that
will not accommodate the larger size.  The pulsar radii are difficult to
infer, and there are very few observational data on the size, but
one estimate of the 4.2 Hz Geminga pulsar's radius concludes that it is
less than 9.5 km \cite{shearer} if the thermal emissivity is 1, which is not
consistent with the oscillatory model.  The estimate is based on an
upper limit for the unpulsed blackbody optical brightness.  No unpulsed
optical emission was detected, and a demonstration that the unpulsed
optical component actually is blackbody would be helpful in establishing
the result.  An even more essential consideration is that if the surface
is highly conductive then it will be specular, which translates into a
low thermal emissivity.

At still higher frequencies there is the constraint that the oscillatory
calculations become so inaccurate as to be unusable when the
diameter is comparable to a quarter wavelength, and, depending on the
value of $b$, that problem can arise with the youngest conventional
pulsars.  The neglect of the magnetic field in the calculations can also
cause the size to be overestimated when the frequency is relatively
high.  The current flow on the surface allows the external magnetic
field to oscillate, but the internal magnetic field of a highly
conductive sphere cannot change quickly.  It may nevertheless be
possible for mechanical deformations to generate internal $E$ fields,
which cannot exist either, but if the $E$ field due to the rate of change
of the magnetic field is equal and opposite to the $E$ field due to
mechanical deformation then the interior magnetic field can oscillate at
the pulsar frequency.  Such an effect, if it exists, would significantly
affect the higher frequency solutions.  The low frequency solutions
would not be much affected.

While the gravitational inductance looks like an inductor at any given
frequency, the frequency dependence is quite different.  In particular,
the equations illustrate that if the vacuum surrounding the pulsar is
replaced by a material with a higher dielectric constant then the
frequency \emph{increases}.  Thus if there are additional interior energy
terms that are in phase with the energy of the $E$ field then the
frequency will increase, which will result in a smaller pulsar for a
given frequency and mass.

There is then the possibility that a more refined derivation (along with
some new theory) will predict smaller young pulsars, and by avoiding the
regime of questionable accuracy at first the equations as given should
be satisfactory for observationally discriminating between the rotary
and oscillatory pulsar models.

Supplemental material, including detailed derivations of all the
above equations, is available on the Internet at www.s-4.com/pulsar.
Section 5 of the web page includes images with the drift lines drawn on them
for some of the drifting subpulse bands analyzed here.



\begin{thebibliography}{}
\bibitem[Baring \& Harding (1998)]{baring}
Baring, M., \& Harding, A. 1998 ApJ, \textbf{507-1}, L55

\bibitem[(Biggs et al 1985)]{biggs}
Biggs, J. D., McCulloch, P. M., Manchester, R. N., Lyne, A. G. 1985,
    MNRAS, \textbf{215}, 281

\bibitem[(Bildsten et al 1997)]{bildsten} Bildsten, L. et al 1997 ApJS,
    \textbf{113}, 367

\bibitem[(Cadez \& Galicic 1996)]{cadez}
Cadez, A., \& Galicic, M. 1996, A\&A, \textbf{306}, 443

\bibitem[(Callanan et al 1998)]{callanan}
Callanan, P., Garnavich, P., Koester, D. 1998, MNRAS, \textbf{298-1}, 207

\bibitem[(Calucci 1999)]{calucci}
Calucci, G. 1999 Modern Physics A, \textbf{14}, L18

\bibitem[(Damour et al 1993)]{damour}
Damour, T., Deser, S., McCarthy J. 1993, Phys.Rev.D \textbf{47}, 1541

\bibitem[(Deshpande \& Rankin)]{deshpande00}
Deshpande, A. A., \& Rankin, J. M., MNRAS, in press (astro-ph/0010048)

\bibitem[(de Felice 1971)]{felice}
de Felice, F. 1971 Gen. Relativ. and Gravit. \textbf{v. 2 no. 4}, 347

\bibitem[(Golden et al 2000)]{golden}
Golden, A., Redfern, R. M., Beskin, G. M., Neizvestny, S. I., Neustroev, V. V.,
Plokhotnichenko, V. L., Cullum, M. 2000, A\&A, \textbf{363}, 617

\bibitem[(Gotthelf et al 1999)]{gotthelf}
Gotthelf, E. V., Vasisht, G., Dotani, T. 1999, ApJ, \textbf{522} Issue 1, L49

\bibitem[(Hankins \& Wolszan 1987)]{hankins}
Hankins, T. H., \& Wolszan, A. 1987, ApJ \textbf{318} part 1, 410

\bibitem[(Helfand \& Gotthelf)]{helfand}
Helfand, D.J., \& Gotthelf, E.V., ApJ, submitted  (astro-ph/0007310)

\bibitem[(Hurley et al 1999)]{hurley}
Hurley, K. et al 1999, Nature, \textbf{397}, 41

\bibitem[(Israel et al 1999)]{israel}
Israel, L., Covino, S., Stella, L., Campana, S., Haberl, F., Mereghetti, S.
1999, ApJ, \textbf{518} Issue 2, L107

\bibitem[(Kouveliotou et al 1998)]{kouveliotou}
Kouveliotou, C. et al 1998, Nature, \textbf{393}, 235

\bibitem[(Marsden et al 1999)]{marsden}
Marsden, D., Rothschild, R. E., Lingenfelter, R. E. 1999
ApJ, \textbf{520} Issue 2,  L107

\bibitem[(Moffat 1995)]{moffat} Moffat, J. W. 1995,
    Phys.Lett.B \textbf{355}, 447

\bibitem[(Morse \& Feshbach 1953)]{morse} P. M.Morse and H. Feshbach, \textit{Methods of Theoretical
    Physics} (McGraw--Hill, New York, 1953) Vol 1

\bibitem[(Pandey 1996)]{pandey}
Pandey, U. S. 1996, A\&A, \textbf{316}, 111

\bibitem[Shabad \& Usov (1982)]{shabad}
Shabad, A. E., \& Usov, V. V. 1982, Nature, \textbf{295}, 215

\bibitem[(Shearer et al 1999)]{shearer}
Shearer, A., Golden, A., O'Conner, P., Beskin, G., Redfern, M. 1999,
Irish Astronomical J., \textbf{26(2)}, 99

\bibitem[(Sieber \& Oster)]{sieber}
Sieber, W., \& Oster, L. 1975, A\&A, vol. \textbf{38} no. 2, 325

\bibitem[(Smith et al 1997)]{smith}
Smith, I. A., Schultz, A. S., Hurley, K., van Paradijs, J., Waters, L. B. 1997,
A\&A \textbf{319}, 923

\bibitem[(Stairs et al 2000)]{stairs}
Stairs, I. H., Lyne, A. G., Shemar, S. L. 2000, Nature, \textbf{406}, 484

\bibitem[(Unwin et al 1978)]{unwin}
Unwin, S. C., Readhead, A.C.S., Wilkinson, P. N., Ewing, W. S. 1978,
MNRAS, \textbf{182}, 711

\bibitem[(Vasisht \& Gotthelf 1997)]{vasisht}
Vasisht, G., \& Gotthelf, E. 1997, ApJ, \textbf{486}, L129

\bibitem[(Vivekanand \& Joshi 1997)]{vivekanand}
Vivekanand, M., \& Joshi, B. C. 1997, ApJ, \textbf{477}, 431

\bibitem[(Weisskopf et al 2000)]{weisskopf}
Weisskopf, M. C. et al 2000, ApJ, \textbf{536} Issue 2, L81

\bibitem[(Woods et al 2000)]{woods}
Woods, P. M., Kouveliotou, C., Finger, M. H., Gogus, E., Swank, J. Smith, D. A.
2000 American Astron. Soc.,  HEAD meeting \#32, \#36.02

\end{thebibliography}
\end{document}